\documentclass[prl,aps,superscriptaddress,twocolumn,letter,nopacs,footinbib,longbibliography]{revtex4-2}
\pdfoutput=1

\usepackage{amsmath}
\usepackage{graphicx}
\usepackage{amsfonts}
\usepackage{amsthm}

\usepackage[hypertexnames=false]{hyperref}

\hypersetup{colorlinks=true,
  linkcolor=blue,
  citecolor=blue,
  urlcolor=blue,
  breaklinks=true,   
  colorlinks=true,   
  pdfusetitle=true  
}

\usepackage[margin=2cm]{geometry}

\newcommand{\ket}[1]{\left\lvert{#1}\right\rangle}
\newcommand{\bra}[1]{\left\langle{#1}\right\rvert}
\newcommand{\braket}[2]{\left\langle{#1}\middle | {#2}\right\rangle}
\newcommand{\ketbra}[2]{\left\lvert{#1}\middle \rangle\middle\langle{#2}\right\rvert}
\newcommand{\norm}[1]{\left\lvert\left\lvert{#1}\right\rvert\right\rvert}
\newcommand{\abs}[1]{\left\lvert{#1}\right\rvert}
\newtheorem{thm}{Theorem}
\newtheorem{defi}{Definition}
\newtheorem{lem}{Lemma}

\begin{document}

\title{The fermionic linear optical extent is multiplicative for $4$ qubit parity eigenstates}
\author{Oliver Reardon-Smith}
\affiliation{Center for Theoretical Physics, Polish Academy of Sciences, Warsaw}

\begin{abstract}
The Fermionic linear optical (FLO) extent is a quantity that serves two roles, firstly it serves as a measure of the ``quantumness'' (or non-classicality) of quantum circuits. Secondly it controls the runtime of a class of classical simulation algorithms, which are state-of-the-art for simulating quantum circuits formed mostly of FLO unitaries and promoted to universality by the addition of ``magic states''. It is therefore interesting to understand the scaling behaviour of the extent as magic states are added to a circuit. In this work we solve this problem for the case of $4$-qubit parity eigenstates. We show that the FLO extent of a tensor product of any pure state and a $4$ qubit parity eigenstate is the product of the extents of the two tensor factors. Applying this result recursively one proves a conjecture that the extent is multiplicative for arbitrary tensor products of $4$ qubit magic states.

\end{abstract}

\maketitle
\renewcommand{\thefootnote}{\arabic{footnote}}
\setcounter{footnote}{0}

\paragraph{Introduction.}

In the last few years work on classical simulation algorithms for quantum circuits has brought attention to a quantity known as the \emph{stabilizer extent}~\cite{bravyi-2019,PRXQuantum.3.020361,Heimendahl2021stabilizerextentis}. While this quantity serves as a measure of the non-classicality\footnote{More specifically non-stabilizerness.} of a quantum computation, of more practical interest is its role as a bound on the exponential component of the runtime of practical classical simulation algorithms~\cite{bravyi-2019,PRXQuantum.3.020361}. These simulation algorithms are state-of-the-art for quantum circuits consisting primarily of Clifford gates, promoted to university by some non-Clifford gates or non-stabilizer input ``magic states''. The exponential behaviour of the runtime of these simulation algorithms follows directly from the fact that the stabilizer extent is multiplicative for tensor products of magic states~\cite{bravyi-2019}. This link may be summarised in the slogan - the more quantum the computation is, the slower a classical simulation of it will be.

Stabilizer quantum mechanics is not the only efficiently classically simulable subtheory of quantum mechanics~\cite{terhal2002Classical,doi:10.1098/rspa.2009.0433,doi:10.1098/rspa.2002.1097}, another well-known example is fermionic linear optics (or FLO, for brevity). FLO circuits are equivalent, under the Jordan-Wigner transformation, to those formed of nearest-neighbour matchgates~\cite{valiant-2001,terhal2002Classical,jordan-wigner-1928}\footnote{In this work we will discuss our results in terms of matchgates for notational clarity however all of our results are valid for any qubit encoding of FLO and are not limited to the Jordan-Wigner transformation.}. In recent years attention has been brought to applications fermionic linear optics including in the context of classical shadows~\cite{matchgate-shadows,error-mitigated-matchgate-shadows}, estimation of fermionic observables~\cite{Fermionic-Joint-Measurements,estimating-fermionic-observables}, and Heisenberg limited learning of fermionic observables~\cite{Learning-interacting-fermionic-Hamiltonians}. They have also been a fruitful source of complexity theoretic results including the quantum advantage scheme of Ref.~\cite{PRXQuantum.3.020328}, and the unexpected result that, in stark contrast to stabilizer states, PAC learning fermionic Gaussian states is NP-hard~\cite{pac-learning-free-fermion}.

In addition to being an central model of quantum computation, fermionic linear optics naturally represents the physics of a system of spinless fermions under a quadratic Hamiltonian. The circuits our results apply to, consisting of mostly FLO gates, with the addition of non-FLO magic naturally represent the evolution of such fermions with limited interaction terms perturbing the free Hamiltionians~\cite{oz2017}. Our results therefore apply to a class of operations which have already been applied in the context of quantum simulations of chemistry and many-body systems~\cite{doi:10.1126/science.abb9811,PhysRevLett.120.110501,PhysRevApplied.9.044036,Dallaire-Demers_2019,Phasecraft2022,UnbiasingMonteCarlo2022,matchgate-shadows}.

These applications may involve the application of FLO or matchgate operations input states which are not themselves fermionic Gaussian states, thus escaping from the efficient simulability results of~\cite{valiant-2001,terhal2002Classical,PhysRevA.93.062332}. This motivates the generalization of the simulation algorithms of the first paragraph from stabilizer quantum mechanics to matchgate quantum mechanics. Recently this goal has been achieved for the subtheory known as fermionic linear optics (FLO)~\cite{Dias_2024,reardonsmith2024improvedsimulationquantumcircuits}. These works directly generalize the results known from the stabilizer case, obtaining classical simulation algorithms for circuits consisting of FLO unitaries promoted to universality by the addition of fixed parity magic states (or, equivalently, controlled rotation gates with arbitrary phases)\cite{Yoganathan-2019-magic-for-matchgates}. The non-polynomial component of the runtimes of these algorithms is comprised of a linear dependence on the \emph{fermionic linear optical extent}. Under the natural conjecture that the extent is multiplicative for magic states, just as the stabilizer extent is, this non-polynomial component is exactly the exponential scaling we expect to obtain when simulating quantum computations with a classical computer.

However, while Refs.~\cite{Dias_2024,reardonsmith2024improvedsimulationquantumcircuits,cudby2024gaussiandecompositionmagicstates} obtain the FLO extent for a single magic state, and show that the extent of a product of two magic states is the product of the individual extents, the behaviour of the extent for products of more magic states is unknown. In this work we remedy this situation and show that the extent of an arbitrary tensor product of $n$ FLO magic states is simply the product of the individual extents of the tensor factors. A proof of this result was announced in Ref.~\cite{cudby2024gaussiandecompositionmagicstates} however it appears to contain an error which renders it incomplete. See Supplemental Material for details.

In fact our main result, theorem~\ref{thm:extent-multiplicative} is slightly stronger than this. We show that the FLO-extent of a tensor product of an arbitrary $n$ qubit pure state and any fixed-parity $4$ qubit state is simply the product of the extents of the tensor factors. While the first side of this, the arbitrary quantum state, is a considerable generalization; the second side, the fixed parity $4$ qubit state is less of a generalization than it would appear since all fixed-parity $4$ qubit states are FLO magic states. It has been noted in~\cite{Dias_2024,reardonsmith2024improvedsimulationquantumcircuits,cudby2024gaussiandecompositionmagicstates} that multiplicativity results for the FLO fidelity imply those for the FLO-extent. Interestingly, however, while our proof of theorem~\ref{thm:extent-multiplicative} does involve a multiplicativity result for the fidelity we do not need anything close to as general as theorem~\ref{thm:extent-multiplicative} for the FLO fidelity. Indeed we are only able to show the fidelity is multiplicative for a specific class of maximally magic $4$ qubit states. 

\paragraph{Setting.} We first briefly introduce Fermionic linear optics from the point of view of the Jordan-Wigner transformation, since this embeds the state-space into that of an $n$-qubit system, and is hence easily understood by quantum information theorists. Label the $n$ qubits with integers from $0$ to $n-1$ and define the \emph{Majorana fermion operators}
\begin{align}
  c_{2j} &= \left(\prod_{k=0}^{j-1} Z_k\right) X_j\\
  c_{2j+1} &= \left(\prod_{k=0}^{j-1} Z_k \right)Y_j,
\end{align}
where $X_i$, $Y_i$ and $Z_i$ are the Pauli operators on the $i^\text{th}$ qubit. The Majorana fermion operators obey the standard anticommutation relation $\{c_i, c_j\} = 2\delta_{ij} I$ and hence define fermionic creation and annihilation operators by the relations
\begin{align}
  c_{2j} &= (a_j + a_j^\dagger)\\
  c_{2j+1} &= -i(a_j - a_j^\dagger).
\end{align}
One identifies the all-zero computational basis state with the fermionic vacuum state, and a computational basis state containing a $\ket{1}$ on any of the qubits with a fermionic state with those modes occupied. A unitary $U$ is called fermionic linear optical (FLO) if it obeys the relation
\begin{align}
  U c_i U^\dagger = \sum_j R_{ij} c_j,\label{eqn:flo-unitary}
\end{align}
for some real matrix $R$, a consequence of~\eqref{eqn:flo-unitary} is that $R$ is orthogonal. Fermionic Gaussian states are just those which are obtained from a computational basis state by applying a FLO unitary, we will also call these FLO states for conciseness, and call the class of $n$ qubit FLO states $\mathrm{FLO}(n)$. Fermionic linear optics is a restricted model of quantum computation, but it may be promoted to universality by the addition of non-FLO \emph{magic-states}~\cite{PhysRevA.73.042313, Yoganathan-2019-magic-for-matchgates}. Since the term ``magic'' is primarily used in Clifford/stabilizer quantum mechanics we will call the magic states we are interested in FLO magic states. From Ref.~\cite{Yoganathan-2019-magic-for-matchgates} we know that FLO magic-states must be \emph{fermionic} or eigenstates of the parity operator. Since all fixed-parity states of one, two and three qubits are FLO states the first interesting case is that of four qubits. We will label the $4$ qubit magic states of Ref.~\cite{Yoganathan-2019-magic-for-matchgates} as
\begin{align}
  \ket{M_\phi} &= \frac{1}{2}\left(\ket{0000}+\ket{0011}+\ket{1100} +e^{i\phi}\ket{1111}\right),
\end{align}
in fact any fixed parity $4$ qubit pure state is proportional to $U\ket{M_\phi}$ for a FLO unitary $U$ and some $\phi\in\mathbb{R}$, while $\ket{M_\phi}$ with different $\phi$ ($\operatorname{mod} 2\pi$) are inequivalent by FLO unitaries. Finally we define the two resource theoretic quantities relevant in this work, are the FLO fidelity and FLO extent, definitions~\ref{def:flo-fidelity}~\ref{def:flo-extent}, respectively.
\begin{defi}[FLO fidelity]\label{def:flo-fidelity}
  For an $n$ qubit quantum state $\ket{\psi}$, the FLO fidelity is
  \begin{align}
    \operatorname{F}_\text{FLO}(\ket{\psi}) &= \max_{\ket{\omega}\in\mathrm{FLO}(n)} \left\lvert\braket{\omega}{\psi}\right\rvert^2.
  \end{align}
\end{defi}
\begin{defi}[FLO extent]\label{def:flo-extent}
  For an $n$ qubit quantum state $\ket{\psi}$, the FLO extent is
  \begin{align}
    \operatorname{\xi}_\text{FLO}(\ket{\psi}) &= \inf\left\{\norm{a}^2_1\middle| \sum_ja_j\ket{\omega_j} = \ket{\psi}\right\},
  \end{align}
  where the states $\ket{\omega_j}$ are FLO states. Note that we do not put a restriction on the dimension of the complex vector $a$, other than that it is finite.
\end{defi}
These two definitions are linked by a very useful characterisation of the extent for general resource theories in terms of the convex dual problem,
\begin{align}
  \operatorname{\xi}_\text{FLO}\left(\ket{A}\right) &= \max_{\ket{\omega}}\frac{\lvert\braket{\omega}{A}\rvert^2}{\operatorname{F}_\text{FLO}(\ket{\omega})},\label{eqn:extent-dual-problem}
\end{align}
where the $\max$ is over all pure states $\ket{\omega}$ of appropriate dimension. See Ref.~\cite{Dias_2024,Regula_2017} or, for an application to the stabilizer extent, Ref.~\cite{bravyi-2019}.

\paragraph{Results.} We begin with what is arguably the main technical result of our work, lemma~\ref{lem:fidelity-fixed-parity}. We leave proofs to the Supplemental Material but give a summary of the idea here. We employ the characterization of entanglement of FLO states from Ref.~\cite{BOTERO200439} to reduce the problem to one of $4$ qubit systems, then apply a decomposition of each space of fixed-parity $4$-qubit states into orbits under FLO unitaries due to Ref.~\cite{Oszmaniec2014}.
\begin{lem}[Multiplicativity of FLO fidelity for fixed parity states]\label{lem:fidelity-fixed-parity}
  Let
  \begin{align}
    \ket{a_8} &= \frac{1}{\sqrt{2}}\left(\ket{0000} + \ket{1111}\right),
  \end{align}
  and let $\ket{\phi}$ be an $n$ qubit state of fixed parity, then
  \begin{align}
    \operatorname{F}_\text{FLO}\left(\ket{\phi}\ket{a_8}\right) &= \operatorname{F}_\text{FLO}\left(\ket{\phi}\right)\operatorname{F}_\text{FLO}\left(\ket{a_8}\right)\\
                                                                &= \operatorname{F}_\text{FLO}\left(\ket{\phi}\right)\frac{1}{2}
  \end{align}
\end{lem}
Here we note a simplification of the convex dual problem for the extent~\eqref{eqn:extent-dual-problem} in the case of fermionic linear optics. 
\begin{lem}[Characterisation of extent]\label{lem:extent-only-needs-one-subspace}
  Let $\mathcal{H}_{\pm}$ be the subspaces of the Hilbert space corresponding to positive and negative eigenvalues of the parity operator. Let $x\in\{+,-\}$ and $\ket{\psi} \in \mathcal{H}_{x}$ then 
  \begin{align}
    \operatorname{\xi}_\text{FLO}\left(\ket{\psi}\right) &= \max_{\ket{\omega}\in \mathcal{H}_{x}}\frac{\lvert\braket{\omega}{\psi}\rvert^2}{\operatorname{F}_\text{FLO}(\ket{\omega})}.\label{eqn:extent-only-needs-one-subspace}
  \end{align}
\end{lem}
Combining Lemmas~\ref{lem:fidelity-fixed-parity} and~\ref{lem:extent-only-needs-one-subspace} with the parametrization of FLO magic states due to~\cite{Oszmaniec2014} one obtains lemma~\ref{lem:extent-multiplicative-fixed-parity}.
\begin{lem}[Mutiplicativity of FLO extent for fixed parity states]\label{lem:extent-multiplicative-fixed-parity}
  Let $\ket{\psi}$ be an $n$ qubit state of fixed parity, and let $\ket{M}$ be an even $4$ qubit state. Then
  \begin{align}
    \operatorname{\xi}_\text{FLO}\left(\ket{\psi}\ket{M}\right) &= \operatorname{\xi}_\text{FLO}\left(\ket{\psi}\right)\operatorname{\xi}_\text{FLO}\left(\ket{M}\right).
  \end{align}
\end{lem}
\begin{thm}[Mutiplicativity of FLO extent]\label{thm:extent-multiplicative}
  Let $\ket{\psi}$ be an $n$ qubit state, and let $\ket{M}$ be an even $4$ qubit state. Then
  \begin{align}
    \operatorname{\xi}_\text{FLO}\left(\ket{\psi}\ket{M}\right) &= \operatorname{\xi}_\text{FLO}\left(\ket{\psi}\right)\operatorname{\xi}_\text{FLO}\left(\ket{M}\right).
  \end{align}
\end{thm}

\paragraph{Outlook} An interesting feature of our proof is that while we rely on a multiplicativity result for the fidelity to prove one for the extent, but the result for the extent we obtain is substantially more general than the one for the fidelity. We are only able to show multiplicativity for the fidelity for a relatively narrow class of states, those equivalent to $\ket{a_8}$ under FLO unitaries. It so happens that this class contains the witness states required for the extent (the optimizing $\ket{\omega}$ in equation~\eqref{eqn:extent-only-needs-one-subspace}). It remains to be seen whether the FLO fidelity is multiplicative for $4$ qubit fermionic states.
While the extent and fidelity are both interesting and practically relevant quantities in their own right, the ultimate theoretical runtime scaling of classical simulation algorithms such as~\cite{PhysRevLett.116.250501,bravyi-2019,PRXQuantum.3.020361,reardonsmith2024improvedsimulationquantumcircuits,Dias_2024} is given by the (approximate) \emph{rank}, for which the extent provides an upper bound. Little is known about the behaviour of the FLO rank for products of FLO magic states, with the investigations in Ref.~\cite{cudby2024gaussiandecompositionmagicstates} being the state-of-the-art. In this work we have employed the results of Ref.~\cite{Oszmaniec2014}, which completely characterize the orbits of $4$ qubit fixed parity states under FLO unitaries. It remains to be seen whether the techniques we have employed here imply relevant constraints on the FLO-rank or approximate rank.

\paragraph{Acknowledgements} O.~R.-S. acknowledges funding from National Science Centre, Poland under the grant OPUS: UMO2020/37/B/ST2/02478.
\clearpage
\bibliography{floextent}
\pagebreak

\setcounter{equation}{0}
\setcounter{figure}{0}
\renewcommand{\theequation}{S\arabic{equation}}
\renewcommand{\thefigure}{S\arabic{figure}}
\onecolumngrid
\widetext
\begin{center}
	\textbf{\large Supplemental Material}
\end{center}

\section{The ``magic basis''}
\label{sec:app-magic-basis}
In this section we will summarize the results of Refs.~\cite{Oszmaniec2014} which are relevant for this work. There is a ``magic basis'' $\{\ket{\eta_i}\}_{i=1}^8$
\begin{align}
  \ket{\eta_1} &= \frac{1}{\sqrt{2}}\left(\ket{0000} + \ket{1111}\right), &
                                                                            \ket{\eta_2} &= \frac{i}{\sqrt{2}}\left(\ket{0000} - \ket{1111}\right)\nonumber\\
  \ket{\eta_3} &= \frac{1}{\sqrt{2}}\left(\ket{0011} - \ket{1100}\right),&
                                                                           \ket{\eta_4} &= \frac{i}{\sqrt{2}}\left(\ket{0011} + \ket{1100}\right)\nonumber\\
  \ket{\eta_5} &= \frac{1}{\sqrt{2}}\left(\ket{0101} + \ket{1010}\right), &
                                                                            \ket{\eta_6} &= \frac{i}{\sqrt{2}}\left(\ket{0101} - \ket{1010}\right)\nonumber\\
  \ket{\eta_7} &= \frac{1}{\sqrt{2}}\left(\ket{1001} - \ket{0110}\right),&
                                                                           \ket{\eta_8} &= \frac{i}{\sqrt{2}}\left(\ket{1001} + \ket{0110}\right)\nonumber,
\end{align}
with the property that for any even, $4$-qubit FLO state $\ket{M}$ there exists real, $8$ dimensional vectors $r$ and $s$ such that $\norm{r}_2 = \norm{s}_2 = 1$, $r\cdot s  = 0$ and
\begin{align}
  \ket{M} &\propto \sum_j \left(\cos(a)r_j + i\sin(a) s_j\right) \ket{\eta_j},\label{eqn:magic-basis-decomp}
\end{align}
for some $a\in (-\pi,\pi]$. We will neglect the $\propto$ and (without loss of generality) assume the overall phase of $\ket{M}$ is such that~\eqref{eqn:magic-basis-decomp} is an equality. Note that the action of $4$ qubit FLO unitaries is transitive on $(r,s)$ pairs, in particular for any $(r,s)$ there exists a FLO unitary $U$ such that
\begin{align}
  U\left(\sum_j \left(\cos(a)r_j + i\sin(a) s_j\right) \ket{\eta_j}\right) &= \cos(a)\ket{\eta_1} + i\sin(a)\ket{\eta_2}.
\end{align}
The parameter $a$ in equation~\eqref{eqn:magic-basis-decomp} may therefore be seen as labelling the orbits of $4$-qubit FLO unitaries, since we may multiply the whole of~\eqref{eqn:magic-basis-decomp} by a factor of $i$ to swap the role of the $\cos$ and $\sin$ and absorb factors of $-1$ into the vectors $r$ and $s$, the orbits are uniquely determined if we restrict to $a\in\left[0,\frac{\pi}{4}\right]$. The decomposition~\eqref{eqn:magic-basis-decomp} for the magic states $\ket{M_\phi}$ takes the form
\begin{align}
\ket{M_\phi} &= \frac{1}{2}\left(\ket{0000} + \ket{0011} + \ket{1100} + e^{i\phi} \ket{1111}\right)\nonumber \\
e^{i\frac{\pi-\phi}{4}} \ket{M_\phi} &= \cos\left(\frac{\phi+\pi}{4}\right)\left( \cos\left(\frac{p}{2}\right)\ket{\eta_1} - \sin\left(\frac{p}{2}\right)\ket{\eta_2} + \ket{\eta_4}\right) +\\&\quad i \sin\left(\frac{\phi+\pi}{4}\right)\left( \cos\left(\frac{p}{2}\right)\ket{\eta_1} - \sin\left(\frac{p}{2}\right)\ket{\eta_2} -\ket{\eta_4}\right),\label{eqn:magic-state-magic-basis-decomp}
\end{align}
i.e.
\begin{align}
\begin{pmatrix}r_1 \\r_2\\r_4\end{pmatrix} &= \frac{1}{\sqrt{2}}\begin{pmatrix}\cos\left(\frac{\phi}{2}\right)\\ -\sin\left(\frac{\phi}{2}\right)\\1\end{pmatrix}\\
\begin{pmatrix}s_1 \\s_2\\s_4\end{pmatrix} &= \frac{1}{\sqrt{2}}\begin{pmatrix}\cos\left(\frac{\phi}{2}\right)\\ -\sin\left(\frac{\phi}{2}\right)\\-1\end{pmatrix},
\end{align}
and the other $5$ components of each vector are $0$.

It is easily verified that the FLO-fidelity of a single $4$ qubit even state is given by the following expression
\begin{align}
  \operatorname{F}_{\text{FLO}}\left(\cos(a)\ket{\eta_1} + i\sin(a)\ket{\eta_2}\right) &= \max_{r,s} \abs{\frac{1}{\sqrt{2}}\sum_j (r_j -i s_j)\bra{\eta_j}\left(\cos(a)\ket{\eta_1} + i\sin(a)\ket{\eta_2}\right)}^2\\
                                                                                       &=  \frac{1}{2}\max_{r,s\in\mathbb{R}^8} (r_1\cos(a) + s_2\sin(a))^2 + (r_2\sin(a) - s_1\cos(a))^2\\
                                                                                       &=\frac{1}{2}(\abs{\cos(a)} + \abs{\sin(a)})^2\\
                                                                                       &=\frac{1}{2}(1 + \abs{\sin(2a)})
\end{align}

\section{Proof of Lemma~\ref{lem:fidelity-fixed-parity}}
\label{sec:app-prood-fidelity-fixed-parity}
\begin{proof}
  We will prove the result assuming the fixed parity state $\ket{A}$ is of even parity. The proof is almost identical for $\ket{A}$ of odd parity so we will omit it. In addition to the results of Ref.~\cite{Oszmaniec2014}, summarized in section~\ref{sec:app-magic-basis} we will need the main result of Ref.~\cite{BOTERO200439}. For any even-parity FLO state $\ket{\psi}$ in a Hilbert space partitioned into subsystems $A\otimes B$ with $n = \dim(A) \leq \dim(B) = m$ , there exist FLO unitary operators $U_A$ and $U_B$ acting on the respective subsystems and angles $\{\theta_j\}_{j=1}^n$ such that
\begin{align}
  \ket{\psi} = U_A\otimes U_B \prod_{j=1}^n \left(\cos(\theta_j)\ket{0}_{A_j}\ket{0}_{B_j}  + \sin(\theta_j)\ket{1}_{A_j}\ket{1}_{B_j} \right) \ket{0}^{\otimes(m-n)}_B,\label{eqn:botero}
\end{align}
We may assume that in addition to being FLO unitaries, the operators $U_A$ and $U_B$ are both \emph{even} operators. A FLO unitary must be either even or odd (but not a mixture of both). First note that if one is odd, then both must be so that the left and right hand sides of~\eqref{eqn:botero} have the same parity, then observe that if both are odd we may factor a single Majorana fermion operator (e.g. $c_1 = a_1 + a_1^\dagger$) out of each to obtain two even operators. Applying these Majorana fermion operators to the states in parentheses in~\eqref{eqn:botero} swaps the role of $\ket{0}_{A_1}$ with $\ket{1}_{A_1}$ and $\ket{0}_{B_1}$ with $\ket{1}_{B_1}$. We then bring the expression back to one of the same form as~\eqref{eqn:botero} by changing $\theta_1$ to swap the $\cos$ with the $\sin$.

It will be convenient for us to define the notation
\begin{align}
  t_\theta(y) &= \prod_{j=1}^n\cos(\theta_i)^{1-y_i}\sin(\theta_i)^{y_i},
\end{align}
for a binary vector $y$. Multiplying out the brackets in equation~\eqref{eqn:botero} we obtain
\begin{align}
  \ket{\psi} = U_A\otimes U_B \sum_{y\in\{0,1\}^n} t_\theta(y)\ket{y}_A\ket{y}_B\ket{0}^{\otimes(m-n)}\label{eqn:botero-v2},
\end{align}
which we may substitute directly into the expression for the fidelity
\begin{align}
  \operatorname{F}_{\text{FLO}}\left(\ket{a_8}\ket{\phi} \right)   &= \max_{U_A, U_B, \theta} \abs{\sum_{y\in\{0,1\}^4}t_\theta(y) \bra{a_8}U_A\ket{y} \bra{\phi}U_B \ket{y}\ket{0}^{\otimes(m-4)}}^2.
\end{align}
At this point it is convenient to consider instead $\sqrt{F_\text{FLO}\left(\ket{a_8}\ket{\psi}\right)}$, we are going to apply the triangle inequality and then H{\"o}lder's inequality, with $p=1$, $q=\infty$
\begin{align}
  \sqrt{F_\text{FLO}\left(\ket{M}\ket{\psi}\right)} &= \max_{U_A, U_B, \theta} \abs{\sum_{y\in\{0,1\}^4}t_\theta(y) \bra{a_8}U_A\ket{y} \bra{\psi}U_B \ket{y}\ket{0}^{\otimes(m-4)}}\\
                                                      &\leq \max_{U_A, U_B, \theta} \sum_{y\in\{0,1\}^4}\abs{t_\theta(y) \bra{a_8}U_A\ket{y} \bra{\psi}U_B \ket{y}\ket{0}^{\otimes(m-4)}}\\
                                                      &\leq \max_{U_A, U_B, \theta} \sum_{y\in\{0,1\}^4}\abs{t_\theta(y) \bra{a_8}U_A\ket{y}} \max_y \abs{\bra{\psi}U_B \ket{y}\ket{0}^{\otimes(m-4)}}\\
                                                      &= \max_{U_A, \theta} \sum_{y\in\{0,1\}^4}\abs{t_\theta(y) \bra{a_8}U_A\ket{y}} \max_{U_B} \abs{\bra{\psi}U_B \ket{0}^{\otimes m}}\\
  &= \max_{U_A, \theta} \sum_{y\in\{0,1\}^4}\abs{t_\theta(y) \bra{a_8}U_A\ket{y}} \sqrt{F_\text{FLO}\left(\ket{\psi}\right)}.
\end{align}
Having recognised the second term as $\sqrt{\operatorname{F}_{\text{FLO}}(\ket{\psi})}$ we seek to bound the first term. We recall that the FLO-fidelity of $a_8$ is $\frac{1}{2}$ so we are attempting to show the first term is less than $\frac{1}{\sqrt{2}}$. We introduce the real, $8$ dimensional, $2$-norm normalized vector $r$ such that and
\begin{align}
  U_A\ket{M} = \sum_j r_j\ket{\eta_j}.
\end{align}
\begin{align}
  \max_{U_A, \theta} \sum_{y\in\{0,1\}^4}\abs{t_\theta(y) \bra{a_8}U_A\ket{y}} = \max_{r, \theta} &\sum_{y\in\{0,1\}^4}\abs{t_\theta(y) \sum_j r_j\braket{\eta_j}{y}}\\
                                                                             = \max_{r, \theta} &\left(\abs{t_\theta(0000)}+\abs{t_\theta(1111)}\right)\frac{1}{\sqrt{2}}\abs{r_1 + ir_2} +\nonumber\\ &\left(\abs{t_\theta(1100)}+\abs{t_\theta(0011)}\right)\frac{1}{\sqrt{2}}\abs{r_3 + ir_4} +\nonumber\\
  & \left(\abs{t_\theta(1010)}+\abs{t_\theta(0101)}\right)\frac{1}{\sqrt{2}}\abs{r_5 + ir_6} +\nonumber\\&\left(\abs{t_\theta(1001)}+\abs{t_\theta(0110)}\right)\frac{1}{\sqrt{2}}\abs{r_7 + ir_8}.\label{eqn:r-theta-form-first-term}
\end{align}
Defining two $4$ dimensional vectors
\begin{align}
  \tau(\theta) &= \begin{pmatrix}\abs{t_\theta(0000)}+\abs{t_\theta(1111)}\\\abs{t_\theta(1100)}+\abs{t_\theta(0011)}\\\abs{t_\theta(1010)}+\abs{t_\theta(0101)}\\\abs{t_\theta(1001)}+\abs{t_\theta(0110)} \end{pmatrix},\label{eqn:tau-defn}\\
  \rho &= \frac{1}{\sqrt{2}}\begin{pmatrix}\abs{r_1 + ir_2}\\\abs{r_3 + ir_4}\\\abs{r_5 + ir_6}\\\abs{r_7 + i r_8}\end{pmatrix},\label{eqn:rho-defn}
\end{align}
we observe that
\begin{align}
  \norm{\rho}_2^2 = \frac{1}{2} \sum_j r_j^2 = \frac{1}{2}\norm{r}_2^2 = \frac{1}{2},
\end{align}
and
\begin{align}
  \norm{\tau(\theta)}_2^2 &= \frac{1}{2}\left(1 + \abs{\sin(2\theta_1)\sin(2\theta_2)\sin(2\theta_3)\sin(2\theta_4)} + \cos(2\theta_1)\cos(2\theta_2)\cos(2\theta_3)\cos(2\theta_4) \right)\label{eqn:expression-for-tau-norm}\\
                          &\leq \max_\theta \frac{1}{2}\left(1 + \sin(2\theta_1)\sin(2\theta_2)\sin(2\theta_3)\sin(2\theta_4) + \cos(2\theta_1)\cos(2\theta_2)\cos(2\theta_3)\cos(2\theta_4) \right)\\
                          &\leq \max_\theta \frac{1}{2}\left(1 + \sin(2\theta_1)\sin(2\theta_2)+ \cos(2\theta_1)\cos(2\theta_2)\right)\\
                          &\leq \max_\theta \frac{1}{2}\left(1 + \cos(2\theta_1 - 2\theta_2)\right)\\
                          &\leq 1.
\end{align}
The expression~\eqref{eqn:expression-for-tau-norm} for $\norm{\tau(\theta)}_2^2$ may be demonstrated by a tedious calculation, while that for $\norm{\rho}_2^2$ is a consequence of the fact that $r$ is a \emph{real} vector so
\begin{align}
  \abs{r_{2j-1} + i r_{2j}}^2 = r_{2j-1}^2 + r_{2j}^2.
\end{align}
Applying H{\"o}lder's inequality to equation~\eqref{eqn:r-theta-form-first-term}, this time with $p=q=2$ we obtain
\begin{align}
  \max_{U_A, \theta} \sum_{y\in\{0,1\}^4}\abs{t_\theta(y) \bra{a_8}U_A\ket{y}} \leq \frac{1}{\sqrt{2}},
\end{align}
this value is achieved by the choice of parameters $\theta_1 = \theta_2 = \theta_3 = \theta_4 = 0$, $U_A = I$. We also note that if $\ket{\omega}\in \mathrm{FLO}$ is the witness state achieving
\begin{align}
  \abs{\braket{\omega}{\psi}}^2 = F_\text{FLO}\left(\ket{\psi}\right),
\end{align}
then we have the obvious witness state $\ket{0}\ket{\omega}$ achieving
\begin{align}
  \abs{\bra{0}\bra{\omega}\ket{a_8}\ket{\psi}}^2 = \frac{1}{2}F_\text{FLO}\left(\ket{\psi}\right),
\end{align}  
\end{proof}

\section{Proof of theorem~\ref{thm:extent-multiplicative}}
\label{sec:app-extent-multiplicative}

We first prove lemma~\ref{lem:extent-only-needs-one-subspace}.
\begin{proof}
  As mentioned in the main text the following characterisation of the extent is well known
    \begin{align}
    \operatorname{\xi}_\text{FLO}\left(\ket{\psi}\right) &= \max_{\ket{\omega}}\frac{\lvert\braket{\omega}{\psi}\rvert^2}{\operatorname{F}_\text{FLO}(\ket{\omega})},
    \end{align}
    where the $\max$ is over all pure states $\ket{\omega}$ of appropriate dimension. Take a witness state $\ket{\omega}$ and split it into even and odd parts
    \begin{align}
      \ket{\omega} &= a_+\ket{\omega_+} + a_-\ket{\omega_-}. 
    \end{align}
    Now because $\ket{\psi} \in \mathcal{H}_{x}$ and FLO states have fixed parity we have
    \begin{align}
    \operatorname{\xi}_\text{FLO}\left(\ket{\psi}\right) &= \max_{\ket{\omega}}\frac{\abs{a_x}^2\lvert\braket{\omega_x}{\psi}\rvert^2}{\max_{y\in\{+1,-1\}} \abs{a_y}^2 \operatorname{F}_\text{FLO}(\ket{\omega_y})}.
    \end{align}
    If we have
    \begin{align}
      \abs{a_x}^2 \operatorname{F}_\text{FLO}(\ket{\omega_x}) \leq \abs{a_{\bar{x}}}^2 \operatorname{F}_\text{FLO}\left(\ket{\omega_{\bar{x}}}\right),
    \end{align}
    where $\bar{x} = -x$, then reducing $\abs{a_{\bar{x}}}^2$ and increasing $\abs{a_x}^2$ to maintain normalisation will only increase the overall $\max$ the above equation. We may therefore assume that the $\max$ in the denominator is given by $\abs{a_x}^2 \operatorname{F}_\text{FLO}(\ket{\omega_x})$ so the overall expression is
    \begin{align}
      \operatorname{\xi}_\text{FLO}\left(\ket{\psi}\right) &= \max_{\ket{\omega}}\frac{\abs{a_x}^2\lvert\braket{\omega_x}{\psi}\rvert^2}{\abs{a_x}^2 \operatorname{F}_\text{FLO}(\ket{\omega_x})}\\                                                          &=\max_{\ket{\omega}\in\mathcal{H}_x}\frac{\lvert\braket{\omega}{\psi}\rvert^2}{\operatorname{F}_\text{FLO}(\ket{\omega})}.
    \end{align}
\end{proof}

The proof of lemma~\ref{lem:extent-multiplicative-fixed-parity} follows easily from the characterization given in lemma~\ref{lem:extent-only-needs-one-subspace}, combined with the decomposition~\eqref{eqn:magic-state-magic-basis-decomp}.
\begin{proof}
  First note that the extent is trivially submultiplicative so
  \begin{align}
    \operatorname{\xi}_\text{FLO}\left(\ket{\psi}\ket{M}\right) &\leq \operatorname{\xi}_\text{FLO}\left(\ket{\psi}\right)\operatorname{\xi}_\text{FLO}\left(\ket{M}\right).
  \end{align} 
  We use the characterization of the extent in terms of the fidelity
  \begin{align}
    \operatorname{\xi}_\text{FLO}\left(\ket{\psi}\ket{M}\right) &= \max_{\ket{\omega}}\frac{\lvert\braket{\omega}{\psi\otimes M}\rvert^2}{\operatorname{F}_\text{FLO}(\ket{\omega})}.
  \end{align}
  Clearly restricting the $\max$ to be over a smaller set of $\omega$ can not increase the value so we obtain the inequality
  \begin{align}
    \operatorname{\xi}_\text{FLO}\left(\ket{\psi}\ket{M}\right) &\geq \max_{\ket{\omega_1}, \ket{\omega_2}}\frac{\lvert\braket{\omega_1}{\psi}\rvert^2\lvert\braket{\omega_2}{M}\rvert^2}{\operatorname{F}_\text{FLO}(\ket{\omega_1}\ket{\omega_2})}\\
    &=\max_{\ket{\omega_1}}\frac{\lvert\braket{\omega_1}{\psi}\rvert^2}{\operatorname{F}_\text{FLO}(\ket{\omega_1})} \max_{\ket{\omega_2}}\frac{\lvert\braket{\omega_2}{M}\rvert^2}{\operatorname{F}_\text{FLO}(\ket{\omega_2})}
  \end{align}
  where $\ket{\omega_1}$ has fixed parity and $\ket{\omega_2}$ is a $4$ qubit even state of minimal FLO fidelity, i.e. $\ket{\omega_2} =\frac{1}{\sqrt{2}} U(\ket{0000} + \ket{1111})$ for some FLO unitary $U$, and the equality is an application of lemma~\ref{lem:fidelity-fixed-parity}. Applying lemma~\ref{lem:extent-only-needs-one-subspace} shows that the first term is just the FLO-extent of $\ket{\psi}$. Recalling the parametrization from equation~\ref{eqn:magic-basis-decomp}, it remains only to show
  \begin{align}
    \operatorname{\xi}_\text{FLO}(\ket{M}) &= \max_{r}\frac{\lvert\sum_jr_j\braket{\eta_j}{M}\rvert^2}{\operatorname{F}_\text{FLO}(\sum_jr_k\ket{\eta_j})}.
  \end{align}
  Without loss of generality assume
  \begin{align}
    \ket{M} &= \cos(a)\ket{\eta_1} + i\sin(a)\ket{\eta_2},
  \end{align}
  then
  \begin{align}
    \max_{\ket{\omega_2}}\frac{\lvert\braket{\omega_2}{M}\rvert^2}{\operatorname{F}_\text{FLO}(\ket{\omega_2})} &= 2 \max_{r}\lvert r_1\cos(a) + ir_2\sin(a) \rvert^2\\
                                           &= 2 \max\left\{\cos^2(a),\, \sin^2(a) \right\}\\
                                           &=1+\abs{\cos(2a)},
  \end{align}
  recalling the decomposition~\eqref{eqn:magic-state-magic-basis-decomp} we have that 
  \begin{align}
    \max_{\ket{\omega_2}}\frac{\lvert\braket{\omega_2}{M}\rvert^2}{\operatorname{F}_\text{FLO}(\ket{\omega_2})} &= 1+\abs{\sin\left(\frac{\phi}{2}\right)}\\
    &=\operatorname{\xi}_\text{FLO}(\ket{M_\phi}).
  \end{align}
\end{proof}

We now have all the ingredients required to prove theorem~\ref{thm:extent-multiplicative}
\begin{proof}
  If $\ket{\psi}$ is a state of mixed parity it has a unique decomposition into orthogonal even and odd parts (this follows, for example, from the eigen-decomposition of the parity operator). Since all FLO states have fixed parity the only possible decompositions of $\ket{\psi}$ in terms of FLO states are formed by independently decomposing the even and odd parts. Let $\ket{\psi} = a\ket{\psi_e} + b\ket{\psi_o}$ be the orthogal decomposition into even and odd parts respectively and let
\begin{align}
  \ket{\psi_e} &= \sum_j a_j\ket{x_j}\\
  \ket{\psi_o} &= \sum_j b_j\ket{y_j},
\end{align}
be the extent optimal decompositions of the two parts. Then
\begin{align}
  \xi\left(\ket{\psi}\right) &= \left(\abs{a}\sum_j\abs{a_j} + \abs{b}\sum_j\abs{b_j}\right)^2\\
                          &= \left(\abs{a}\sum_j\abs{a_j}+ \abs{b}\sum_j\abs{b_j}\right)^2\\
                          &= \left(\abs{a}\sqrt{\xi\left(\ket{\psi_e}\right)} + \abs{b}\sqrt{\xi\left(\ket{\psi_o}\right)}\right)^2.
\end{align}
Since $\ket{M}$ has fixed parity the parity-eigenspace decomposition of $\ket{\psi}\otimes\ket{M}$ is given by that of $\ket{\psi}$
\begin{align}
  \ket{\psi}\ket{M} = a\ket{\psi_e}\ket{M} + b\ket{\psi_o}\ket{M},
\end{align}
therefore
\begin{align}
  \xi\left(\ket{\psi}\ket{M}\right) &= \xi\left(a\ket{\psi_e}\ket{M} + b\ket{\psi_o}\ket{M} \right)\\
                                 &= \left(\abs{a}\sqrt{\xi\left(\ket{\psi_e}\ket{M}\right)} + \abs{b}\sqrt{\xi\left(\ket{\psi_o}\ket{M}\right)}\right)^2\\
                                 &= \left(\abs{a}\sqrt{\xi\left(\ket{\psi_e}\right)\xi\left(\ket{M}\right)} + \abs{b}\sqrt{\xi\left(\ket{\psi_o}\right)\xi\left(\ket{M}\right)}\right)^2\\
                                 &= \left(\abs{a}\sqrt{\xi\left(\ket{\psi_e}\right)} + \abs{b}\sqrt{\xi\left(\ket{\psi_o}\right)}\right)^2\xi(\ket{M})\\
                                 &= \xi\left(\ket{\psi}\right)\xi(\ket{M}).
\end{align}

\end{proof}

\begin{proof}
  If $\ket{\psi}$ is a state of mixed parity it has a unique decomposition into orthogonal even and odd parts (this follows, for example, from the eigen-decomposition of the parity operator). Since all FLO states have fixed parity the only possible decompositions of $\ket{\psi}$ in terms of FLO states are formed by independently decomposing the even and odd parts. Let $\ket{\psi} = a\ket{\psi_e} + b\ket{\psi_o}$ be the orthogonal decomposition into even and odd parts respectively and let
\begin{align}
  \ket{\psi_e} &= \sum_j a_j\ket{x_j}\\
  \ket{\psi_o} &= \sum_j b_j\ket{y_j},
\end{align}
be the extent optimal decompositions of the two parts. Then
\begin{align}
  \xi\left(\ket{\psi}\right) &= \left(\abs{a}\sum_j\abs{a_j} + \abs{b}\sum_j\abs{b_j}\right)^2\\
                          &= \left(\abs{a}\sum_j\abs{a_j}+ \abs{b}\sum_j\abs{b_j}\right)^2\\
                          &= \left(\abs{a}\sqrt{\xi\left(\ket{\psi_e}\right)} + \abs{b}\sqrt{\xi\left(\ket{\psi_o}\right)}\right)^2.
\end{align}
Since $\ket{M}$ has positive parity the parity-eigenspace decomposition of $\ket{\psi}\otimes\ket{M}$ is given by that of $\ket{\psi}$
\begin{align}
  \ket{\psi}\ket{M} = a\ket{\psi_e}\ket{M} + b\ket{\psi_o}\ket{M},
\end{align}
therefore
\begin{align}
  \xi\left(\ket{\psi}\ket{M}\right) &= \xi\left(a\ket{\psi_e}\ket{M} + b\ket{\psi_o}\ket{M} \right)\\
                                 &= \left(\abs{a}\sqrt{\xi\left(\ket{\psi_e}\ket{M}\right)} + \abs{b}\sqrt{\xi\left(\ket{\psi_o}\ket{M}\right)}\right)^2\\
                                 &= \left(\abs{a}\sqrt{\xi\left(\ket{\psi_e}\right)\xi\left(\ket{M}\right)} + \abs{b}\sqrt{\xi\left(\ket{\psi_o}\right)\xi\left(\ket{M}\right)}\right)^2\\
                                 &= \left(\abs{a}\sqrt{\xi\left(\ket{\psi_e}\right)} + \abs{b}\sqrt{\xi\left(\ket{\psi_o}\right)}\right)^2\xi(\ket{M})\\
                                 &= \xi\left(\ket{\psi}\right)\xi(\ket{M}).
\end{align}

\end{proof}

\section{The apparent issue in the proof of \texorpdfstring{Ref.~\cite{cudby2024gaussiandecompositionmagicstates}}{Cudby and Strelchuk 2023}}

The approach of Ref.~\cite{cudby2024gaussiandecompositionmagicstates} is to attempt to cast the problem of computing the FLO fidelity of a product of un-normalized magic states
\begin{align}
  \ket{M_4}^{\otimes k} &= \left(\ket{0000} + \ket{1111}\right)^{\otimes k},
\end{align}
as a semidefinite program, and use the powerful duality theory of such programs. Unfortunately this is complicated for the following reason. The semidefinite program must include two sets of constraints, the state being optimized over must first be a valid density operator (i.e. both positive-semidefinite and normalized) and secondly must be a fermionic Gaussian state. The first set of constraints is easily expressed as a semidefinite program in the density operator
\begin{align}
  \rho = \ketbra{\psi}{\psi},
\end{align}
while the second set of constraints is instead expressed as a homogeneous quadratic in the computational-basis coefficients of $\ket{\psi}$, i.e. linear constraints in the operator
\begin{align}
  Z_{ij} = \braket{i}{\psi}\braket{j}{\psi}.
\end{align}
Since
\begin{align}
  \rho_{ij} = \braket{i}{\psi}\braket{\psi}{j} \neq Z_{ij},
\end{align}
there is not an obvious way to express the two sets of constraints simultaneously in a semidefinite program. In Ref.~\cite{cudby2024gaussiandecompositionmagicstates} appendix~C. There appears to be an error where the complex conjugate which is the difference between $Z_{ij}$ and $\rho_{ij}$ is lost, specifically in the currently version of the manuscript (v3) on arXiv, the semidefinite program given in equation~(51) is incorrect.

\end{document}